\documentstyle[12pt,axodraw,epsf,epsfig]{article}
\advance\textheight by \topskip
\newcommand{\sla}{\kern -5.4pt /}
\newcommand{\Dir}{\kern -6.4pt\Big{/}}
\newcommand{\Dirin}{\kern -10.4pt\Big{/}\kern 4.4pt}
\newcommand{\DDir}{\kern -7.6pt\Big{/}}
\newcommand{\DGir}{\kern -6.0pt\Big{/}}

\newcommand{\ra}{\rightarrow}
\newcommand{\be}{\begin{equation}}
\newcommand{\ee}{\end{equation}}
\newcommand{\bea}{\begin{eqnarray}}
\newcommand{\eea}{\end{eqnarray}}
\newcommand{\beanon}{\begin{eqnarray*}}
\newcommand{\eeanon}{\end{eqnarray*}}
\newcommand{\ba}{\begin{array}}
\newcommand{\ea}{\end{array}}
\newcommand{\bi}{\begin{itemize}}
\newcommand{\ei}{\end{itemize}}
\newcommand{\ben}{\begin{enumerate}}
\newcommand{\een}{\end{enumerate}}
\newcommand{\bc}{\begin{center}}
\newcommand{\ec}{\end{center}}

\newcommand{\bfig}{\begin{center}\begin{picture}}
\newcommand{\efig}[1]{\end{picture}\\{\small #1}\end{center}}
\newcommand{\flin}[2]{\ArrowLine(#1)(#2)}
\newcommand{\wlin}[2]{\DashLine(#1)(#2){2.5}}

\newcommand{\glin}[3]{\Photon(#1)(#2){2}{#3}}

\newcommand{\sof}{\SetOffset}
\newcommand{\bmip}[2]{\begin{minipage}[t]{#1pt}\bfig(#1,#2)}
\newcommand{\emip}[1]{\efig{#1}\end{minipage}}

\newcommand{\ord}{{\cal O}}

\newcommand{\NP}[1]{{\it Nucl.\ Phys.\ }{\bf #1}}
\newcommand{\PL}[1]{{\it Phys.\ Lett.\ }{\bf #1}}
\newcommand{\ZP}[1]{{\it Z.\ Phys.\ }{\bf #1}}

\newcommand{\JCP}[1]{{\it Jour. Comp. Phys.\ }{\bf #1}}
\newcommand{\CPC}[1]{{\it Comp. Phys. Comm.\ }{\bf #1}}

\begin{document}
\tolerance=100000
\thispagestyle{empty}
\setcounter{page}{0}

\begin{flushright}
{\large DFTT 47/97}\\ 
{\large  PSI--PR--97--18}\\
{\rm October 1997\hspace*{.5 truecm}}\\ 
hep-ph/9710375
\end{flushright}

\vspace*{\fill}

\bc     
{\Large \bf QCD corrections to
$e^+ e^- \to u~\bar d~s~\bar c$ at Lep 2 and the
Next Linear Collider: CC11 at ${\cal O}(\alpha_s )$. 
\footnote{ Work supported in part by Ministero 
dell' Universit\`a e della Ricerca Scientifica.\\[2 mm]
e-mail: maina@to.infn.it, pittau@psw218.psi.ch, pizzio@to.infn.it}}\\[2.cm]
{\large Ezio Maina$^{a}$, Roberto Pittau$^{b}$ and Marco Pizzio$^{a}$}\\[.3 cm]
{\it $^{a}$Dip. di Fisica Teorica, Universit\`a di Torino
     and INFN, Sezione di Torino,}\\
{\it v. Giuria 1, 10125 Torino, Italy.}\\[5 mm]
{\it$^{b}$Paul Scherrer Institute}\\
{\it CH-5232 Villigen-PSI, Switzerland}

\ec

\vspace*{\fill}

\begin{abstract}
{\normalsize
\noindent
QCD one-loop corrections to the full gauge invariant set of electroweak diagrams
describing the hadronic process $e^+ e^- \to u~\bar d~s~\bar c$ are computed.
Four--jet shape variables for $WW$ events are studied at next--to--leading order
and the effects of QCD corrections on the determination of the $W$--mass in the
hadronic channel at Lep 2 and NLC is discussed.
We compare the exact calculation with a ``naive''approach to strong 
radiative corrections which has been widely used in the literature.
}
\end{abstract}

\vspace*{\fill}
\newpage
\section*{Introduction}
The measurement of $W$--mass to high precision is one of the main goals of Lep 2
and will provide a stringent test of the Standard Model (SM) \cite{lep2,Wmass}.
In fact, the mass of the $W$ boson in the SM is tightly constrained and 
an indirect determination of $M_W$
can be obtained from a global fit of all electroweak data.
The fit gives \cite{EWgroupsummer96}
$M_W = 80.338 \pm 0.040^{+0.009}_{-0.018} {\rm GeV}$
where the central value corresponds to $M_H = 300$ GeV and
the second error reflects the change of $M_W$ when the Higgs mass
is varied between 60 and 1000 GeV. 
A disagreement between the value of $M_W$ derived from the global fit and the
value extracted from direct measurement would represent a major failure of the
SM. Alternatively, an improvement in the value of the $W$--mass can 
significantly tighten present bounds on the Higgs mass.
These studies will be continued and improved at the Next Linear Collider (NLC)
which is expected to reduce the error on the measurement of $W$--mass down to
about 15 MeV \cite{ecfa-desy}.

In order to extract the desired information from $WW$ production data,
theoretical predictions with uncertainties smaller than those which are foreseen
in the experiments are necessary. This requires a careful study of all radiative
corrections which have to be brought under control. 
In this paper we will be concerned with QCD corrections which are known
\cite{MP,CC10} to modify the shape of the $W$--mass peak and the distributions
of others kinematical variables. Since event shape variables are often used to
extract $WW$ production from the background it is highly desirable to have a
complete next--to--leading order (NLO)
study of these distributions for $WW$ events. Furthermore,
calculations of QCD corrections to $\ord (\alpha^2 \alpha_s^2)$ four--jet
production have recently appeared \cite{4jNLO}. Combining these results with a
complete
calculation of QCD corrections to all $\ord (\alpha^4)$ four--fermion processes 
it would be possible to obtain NLO predictions for any four--jet shape variable at
Lep~2 and NLC providing new means of testing perturbative QCD.

$W$--pair production can result in a variety of four--fermion hadronic final
states. The simplest one is $u~\bar d~s~\bar c$.
A gauge invariant description of $e^+ e^- \to u~\bar d~s~\bar c$
requires in the unitary gauge the eleven diagrams shown in fig. 1.
This amplitude is known as CC11 in the literature. The subset of three diagrams
labeled (e) and (f) in fig. 1, in which both intermediate
$W$'s can go on mass--shell,
is known as CC03 and is often used for quick estimates of $WW$ production.
Numerically the
total cross sections obtained from CC03 and those obtained from CC11 
at Lep 2 and NLC energies differ by a few per mil \cite{WMC}.
Other final states to which $W$--pairs can decay are $u~\bar u~d~\bar d$ 
($c~\bar c~s~\bar s$) and, through CKM mixing, 
$u~\bar u~s~\bar s$ ($u~\bar u~b~\bar b$, $c~\bar c~d~\bar d$, 
$c~\bar c~b~\bar b$). The additional diagrams which appear in these latter sets
contribute at most at the per--mil level to the total cross section and tend to
produce events which resemble very little $WW$ or even single--$W$ events since 
only electroweak neutral virtual vector bosons appear in them.
There are also four--quark final states where only
neutral intemediate vector boson, including gluons, 
play a role like $u~\bar u~c~\bar c$,
$d~\bar d~s~\bar s$ ($d~\bar d~b~\bar b$, $s~\bar s~b~\bar b$), 
$u~\bar u~u~\bar u$ ($c~\bar c~c~\bar c$) and $d~\bar d~d~\bar d$ ($s~\bar
s~s~\bar s$, $b~\bar b~b~\bar b$). An additional important 
contribution to four--jet production, indistinguishable from four--quark events,
originates from $e^+ e^- \to q~\bar q~g~g$.

In this letter we present the complete calculation of QCD corrections to CC11.
While this is only a first step in the calculation of all $\ord (\alpha^4)$
four--fermion processes at NLO, this reaction includes
the most important source of
background, namely single $W$ production, providing a natural setting for a
first study of the role of QCD corrections to gauge invariant sets of
four--quark production diagrams.

In most instances QCD corrections
 have been included ``naively'' with the substitution 
$\Gamma_W \ra \Gamma_W (1 +2/3\,\alpha_s/\pi)$ and multiplying the hadronic
branching ratio by $(1 +\alpha_s/\pi)$. This prescription is exact for CC03 when
fully inclusive quantities are computed. However, it can only be taken as 
an order of magnitude estimate even for CC03 in the presence of cuts on the jet
directions and properties, as discussed in \cite{MP}.
It is well known that differential distributions can be more sensitive to 
higher order corrections than total cross--sections in which virtual and real
contributions tend to cancel to a large degree.
It is therefore necessary to include higher order QCD effects into the
predictions for $WW$ production and decay in a way which allows to impose
realistic cuts on the structure of the observed events.
The impact of QCD corrections on the angular distribution of the
decay products of a $W$ and their application to on--shell $W$--pair production
is discussed in ref. \cite{lampe}.

\section*{Calculation}
One-loop virtual QCD corrections to 
$e^+ e^- \to u~\bar d~s~\bar c$
are obtained by dressing all diagrams in fig. 1 with gluon loops.  
Defining suitable combinations of diagrams
one can organize all contributions in a very modular way \cite{CC10}.
All QCD virtual corrections to $\ord
(\alpha^4)$ four--fermion processes can be computed using the resulting
set of loop diagrams.

\bfig(160,245)
\sof(-140,140)
\flin{45,5}{65,25}\flin{65,25}{45,45}
\Text(42,5)[rt]{$e^-$}
\Text(42,45)[rb]{$e^+$}
\glin{65,25}{85,25}{4}
\Text(75,30)[b]{$Z, \gamma$}
\sof(-155,140)
\flin{145,45}{125,65}\flin{125,65}{145,85}
\wlin{113.5,38.5}{125,65}
\flin{120,5}{100,25}\flin{100,25}{120,45}
\Text(149,85)[lb]{$s$}
\Text(149,45)[lt]{$\bar c$}
\Text(124,45)[lb]{$u$}
\Text(124,5)[lt]{$\bar d$}
\Text(90,0)[t]{(a)}
\sof(-10,140)
\flin{45,5}{65,25}\flin{65,25}{45,45}
\Text(42,5)[rt]{$e^-$}
\Text(42,45)[rb]{$e^+$}
\glin{65,25}{85,25}{4}
\Text(75,30)[b]{$Z, \gamma$}
\sof(-25,140)
\flin{145,45}{125,65}\flin{125,65}{145,85}
\wlin{113.5,38.5}{125,65}
\flin{120,5}{100,25}\flin{100,25}{120,45}
\Text(149,85)[lb]{$u$}
\Text(149,45)[lt]{$\bar d$}
\Text(124,45)[lt]{$s$}
\Text(124,5)[lt]{$\bar c$}
\Text(90,0)[t]{(b)}
\sof(120,180)
\flin{45,5}{65,25}\flin{65,25}{45,45}
\Text(42,5)[rt]{$e^-$}
\Text(42,45)[rb]{$e^+$}
\glin{65,25}{85,25}{4}
\Text(75,30)[b]{$Z,\gamma$}
\sof(105,180)
\flin{120,5}{100,25}\flin{100,25}{120,45}
\wlin{113.5,11.5}{125,-15}
\flin{145,-35}{125,-15}\flin{125,-15}{145,5}
\Text(124,45)[lb]{$s$}
\Text(124,5)[lb]{$\bar c$}
\Text(149,5)[lb]{$u$}
\Text(149,-35)[lt]{$\bar d$}
\Text(90,-40)[t]{(c)}
\sof(-140,60)
\flin{45,5}{65,25}\flin{65,25}{45,45}
\Text(42,5)[rt]{$e^-$}
\Text(42,45)[rb]{$e^+$}
\glin{65,25}{85,25}{4}
\Text(75,30)[b]{$Z,\gamma$}
\sof(-155,60)
\flin{120,5}{100,25}\flin{100,25}{120,45}
\wlin{113.5,11.5}{125,-15}
\flin{145,-35}{125,-15}\flin{125,-15}{145,5}
\Text(124,45)[lb]{$u$}
\Text(124,5)[lb]{$\bar d$}
\Text(149,5)[lb]{$s$}
\Text(149,-35)[lt]{$\bar c$}
\Text(90,-40)[t]{(d)}
\sof(-10,40)
\flin{45,5}{65,25}\flin{65,25}{45,45}
\Text(42,5)[rt]{$e^-$}
\Text(42,45)[rb]{$e^+$}
\glin{65,25}{85,25}{4}
\Text(75,30)[b]{$Z,\gamma$}
\wlin{85,25}{97,45}
\wlin{85,25}{97,5}
\flin{117,32}{97,45}\flin{97,45}{117,65}
\flin{117,-15}{97,5}\flin{97,5}{117,18}
\Text(121,65)[lb]{$s$}
\Text(121,32)[l]{$\bar c$}
\Text(121,18)[l]{$u$}
\Text(121,-15)[lt]{$\bar d$}
\Text(75,-20)[t]{(e)}
\sof(120,60)
\flin{65,-15}{65,25}\flin{65,25}{45,45}
\flin{45,-35}{65,-15}
\Text(42,-35)[rt]{$e^-$}
\Text(42,45)[rb]{$e^+$}
\wlin{65,25}{95.5,25}
\wlin{65,-15}{95.5,-15}
\flin{110.5,13}{95.5,25}\flin{95.5,25}{110.5,45}
\flin{110.5,-35}{95.5,-15}\flin{95.5,-15}{110.5,-3}
\Text(114.5,45)[lb]{$u$}
\Text(114.5,13)[l]{$\bar d$}
\Text(114.5,-3)[l]{$s$}
\Text(114.5,-35)[lt]{$\bar c$}
\Text(75,-40)[t]{(f)}
\efig{Figure 1: { Tree level diagrams for 
$e^+ e^- \to u~\bar d~s~\bar c$. The dashed lines are $W$'s.}}

The real emission  contribution for $e^+ e^- \to u~\bar d~s~\bar c$
can be obtained attaching a gluon to the quark lines of the diagrams shown in
fig. 1 in all possible positions. This results in fifty--two diagrams.
The required matrix element has been
computed using the formalism presented in ref. \cite{method} with the help of a
set of routines (PHACT) \cite{phact} which generate the building blocks of the
helicity amplitudes  semi-automatically.

The calculation of the virtual corrections has been performed in two different
ways, with identical results. In the first case we have used the standard 
Passarino--Veltman \cite{PV}
reduction procedure, while in the second we have used the new techniques
presented in \cite{red}.
The matching between real and virtual corrections
has been implemented as in \cite{CC10}, using the dipole
formul\ae\ of ref. \cite{catani}.
All integrations have been carried out using the Monte Carlo routine VEGAS 
\cite{vegas}.

An important ingredient for accurate predictions of $W$--pair production is the
effect of initial state radiation (ISR). In the absence of a calculation of all 
${\cal O}(\alpha )$ corrections to four--fermion processes, these effects can
only be included partially. 
In contrast with Lep~1 physics a gauge invariant separation of initial and final
state radiation is not possible. Only the leading logarithmic part of ISR
is gauge invariant and universal. These contributions can be 
included using structure functions. Part of the non--logarithmic terms have been 
computed for CC03 and some other final states in \cite{QEDcorr} using 
the {\it current splitting technique} \cite{split}. 
This corresponds to splitting the electrically neutral 
{\it t}--channel neutrino flow into two oppositely flowing charges, assigning
the $ - 1$ charge to the initial state and the $ + 1$ charge to the final state.
In this way a gauge--invariant definition of ISR can be given. However, there
are clearly cancellations between initial and final state radiation whose
relevance is difficult to estimate in the absence of a complete calculation.
Quite recently the full calculation of all ${\cal O}(\alpha )$ corrections
to $e^+e^- \to  W^+\mu \bar{\nu}_{\mu}$ has been published \cite{QEDtoCC10}.
Non factorizable QED corrections to CC03 have been studied in \cite{nonfact}.
We have only included the leading logarithmic part of ISR using
the $\beta$ prescription in the structure functions, where
$\beta = \ln(s/m^2) - 1$. 
Beamstrahlung effects have been ignored.
We have not included Coulomb
corrections to CC03, which are known to have a sizable effect, particularly
at threshold. They could however be introduced with minimal effort.

\section*{Results}
In this section we present a number of cross sections and distributions
for $e^+ e^- \to u~\bar d~s~\bar c$.  
We have used  $\alpha_s = .123$ at all energies.
ISR is included in all results. The width of the $W$--boson
is kept fixed and includes $\ord (\alpha_s )$ corrections.

For the Lep 2 workshop \cite{lep2}
the so called ADLO/TH set of cuts have been agreed on:
\bi
\item{} the energy of a jet must be greater than 3 GeV;
\item{} two jets are resolved if their invariant mass is larger than 5 GeV;
\item{} jets can be detected in the whole solid angle.
\ei

For the NLC a slightly different set called NLC/TH has been chosen.
The NLC/TH set of cuts differs from the ADLO/TH
set in that a minimum angle of $5^\circ$ is required between a jet and either
beam (as appropriate for the larger bunch size at the NLC and the corresponding
larger bunch disruption
at crossing) and that two jets are resolved if their invariant mass is larger
than 10 GeV. Both set of cuts will also be referred to as ``canonical'' in the
following. We have preferred a different criterion for defining jets which is
closer to the actual practice of the experimental collaborations. For mass
reconstruction studies we have used the Durham scheme \cite{durham}, 
with $y_D = 1.\times 10^{-2}$ at Lep 2 energies. At the NLC we have adopted
a smaller cut $y_D = 1.\times 10^{-3}$ in order to have an adequate fraction of
events with at least four jets.
The four--momenta of the particles which have to be recombined have been simply
summed.
If any surviving jet had an energy smaller than 3 GeV
it was merged with the jet closest in the Durham metric. 

Previous studies \cite{WMC} have shown that the differences between the total 
cross sections
obtained from CC11 and those obtained with CC03 are at the per mil level.
Much larger effects have been found in observables like the average shift of the
mass reconstructed from the decay products from the true $W$--mass. 

In fig. 2 we compare the NLO spectrum of the average reconstructed $W$--mass 
with the naive--QCD (nQCD) result at Lep 2 energies.
All events with at least four observed jets have been retained in fig. 2.
The two candidate masses are obtained forcing first all remaining five--jet 
events to four jets, merging the two partons which are closest in
the Durham scheme, and then selecting the two pairs which minimize
\be\label{selection}
\Delta^\prime_M = \left( M_{R1} 
- M_W \right)^2 + \left( M_{R2} -M_W \right)^2.
\ee
where $M_{R1} and M_{R2}$ are the two candidate reconstructed masses and $M_W$ is
the input $W$--mass.

\begin{figure}
\bc
\begin{tabular}{|c|c|c|c|} \hline 

  & $E_j > 3$ GeV, $y_D = 0.01$ 
  & $E_j > 3$ GeV, $y_D = 0.01$,
  & $E_j > 3$ GeV, $y_D = 0.01$,\\
  & 
  &  $\mid M_{Ri} -M_W \mid < 10$ GeV
  &   $\mid M_{Ri} -M_W \mid < 10$ GeV,\\ 
  &
  &
  & smeared
  \\ \hline\hline
 \rule[-7 pt]{0 pt}{24 pt}
NLO & 1.1493(4) & 0.7895(5) & 0.7758(9)\\ 
\hline
 \rule[-7 pt]{0 pt}{24 pt}
nQCD & 1.1069(3) & 1.0545(3) &1.0479(3) \\ 
\hline
\end{tabular}
\ec
\bc
Table I: Cross sections in pb at $\sqrt{s} = 175$ GeV.
\ec
\end{figure}

In fig. 2 the dashed line refers to the nQCD results while those of the
full NLO calculation are given by the continous line. The corresponding cross
section are given in table I.
Fig. 2a is obtained using only the basic set of cuts described above.
Since all experiments restrict their analysis to a region around the expected
$W$--mass, we have studied the effect of requiring that both reconstructed masses
lie within 10 GeV of the input mass. The result is shown in fig. 2b.
Finally we have tried to take into account some form of experimental smearing,
in order to determine whether the distortion of the mass distribution we observe
does survive in a more realistic setting. To this aim we have smeared the 
reconstructed masses entering eq. (\ref{selection}),
using a gaussian distribution with a 2 GeV width. This procedure gives fig. 2c.
It is worth mentioning that simulating experimental smearing at NLO is far more
complicated than at tree level. In order to preserve the delicate cancellations between
the real emission cross section and the subtraction terms, both contributions to
four--jet quantities must be smeared, event by event, by the same amount.
Fig. 2 shows that at NLO the mass distribution is shifted towards lower
masses and a long tail for rather small average masses is generated, with a
corresponding reduction of the high--mass part of the histogram. This tail
is eliminated when only reconstructed masses in the vicinity of the expected
$W$--mass are retained. Even with this additional cut, however, the NLO
distribution is clearly different from the nQCD one.
The reduction in cross section (see Table I) shows that a large number of soft
gluons is exchanged, at the perturbative level, between decay products of
different $W$'s.
In our simplified treatment
of experimental uncertainties these differences are still visible though
somewhat reduced.
If we try to quantify the mass shift using the standard quantity:
\be
\langle \Delta M \rangle = \frac{1}{\sigma} \int 
\left( \frac{M_{R1} + M_{R2} - 2 M_W }{2} \right) d\sigma.
\ee
we obtain $\langle \Delta M\rangle_{NLO} = -0.229(1)$ GeV and 
$\langle \Delta M\rangle_{nQCD} = -0.0635(4)$ GeV when both reconstructed
masses are required to lie within 10 GeV of the input mass $M_W = 80.23$ GeV
and no smearing is applied.

In fig. 3 we compare the NLO spectrum of the average reconstructed $W$--mass 
with the naive--QCD (nQCD) result at the NLC, using the procedure already
described for the Lep 2 case but for a smaller minimum Durham cut 
$y_D = 1.\times 10^{-3}$. 
Because of the larger relative momentum of the two $W$'s it is less likely that
partons from
the decay of one $W$ end up close to the decay products of the other $W$--boson,
therefore the difference between the two distributions
is smaller than at Lep 2 energies.

In fig. 4 we present the distributions at $\sqrt{s} = 175$ GeV of the
following four--jet shape variables \cite{shape}:
\bi
\item{} the Bengtsson--Zerwas angle: 
 $\chi_{BZ} = \angle [({\bf p_1 \times p_2}),({\bf p_3 \times p_4})]$ (fig. 4a);
\item{} the K\"orner--Schierholz--Willrodt angle :\hfil\break
$\Phi_{KSW} = 1/2 \{\angle [({\bf p_1 \times p_4}),({\bf p_2 \times p_3})]
 + \angle [({\bf p_1 \times p_3}),({\bf p_2 \times p_4})]\}$ (fig. 4b);
\item{} the angle between the two least energetic jets;
$\alpha_{34} = \angle [{\bf p_3},{\bf p_4}]$ (fig. 4c);
\item{} the (modified) Nachtmann--Reiter angle:
$\theta^\ast_{NR} = \angle [({\bf p_1 - p_2}),({\bf p_3 - p_4})]$ (fig. 4d).
\ei
The numbering $i = 1 \dots , 4$ of the jet momenta ${\bf p}_i$ corresponds to
energy--ordered four--jet configurations $( E_1 > E_2 > E_3 > E_4 )$.
We compare the exact NLO results with the distributions
obtained in nQCD and with the results obtained at tree level from the
standard background reactions  $e^+ e^- \to q~\bar q~g~g$, which is the dominant
contribution, and $e^+ e^- \to q_1~\bar {q_1}~q_2~\bar {q_2}$.
 
In all subplots of fig. 4 the full NLO results are given by the continous line
and the nQCD prediction is given by the dashed line. The $q~\bar q~g~g$
and $ q_1~\bar {q_1}~q_2~\bar {q_2}$ tree level background distributions are
given by the chain--dotted and the dotted line respectively. 
The shape variables are computed following the procedure outlined in ref.
\cite{aleph} where the Durham cluster algorithm is complemented by the E0
recombination scheme, namely if the two particles $i$ and $j$ are merged the
pseudo--particle which takes their place remains massless, with four--momentum:
\be
E_{new} = E_i + E_j, \hspace{2 cm} 
{\bf p}_{new}=\frac{E_i + E_j}{|{\bf p}_i +{\bf p}_j |} ({\bf p}_i +{\bf p}_j).
\ee
Using this clustering procedure
all five--jet events are converted into four--jet events,
then each event is used in the analysis if $\min _{i,j=1,4} y_{ij} > y_{cut}$
with $y_{cut} = 0.008$.

In \cite{aleph} it has been shown
that standard parton shower Monte Carlo programs like JETSET do not reproduce
well the observed distribution of four--jet shape variables at Lep,
which are instead
well described by HERWIG 5.9{\footnotesize A} which includes four parton
matrix elements.
The discrepancy can be partially explained by the different distributions
for the $q~\bar q~g~g$ and the $ q_1~\bar {q_1}~q_2~\bar {q_2}$ final states,
which can be clearly seen in fig. 4, since the four quark final state is
included in the parton shower programs only partially.
It is precisely in correspondence with the peaks of the $ q_1~\bar
{q_1}~q_2~\bar {q_2}$ distributions that the discrepancy between data and
simulations is larger.

From fig. 4 it is apparent that at NLO four--jet shape variables distributions
are significantly modified with respect to leading order results, which are
indistinguishable from the nQCD distributions.
It is also evident that four--jet distributions in $WW$ events are markedly
different from the background distributions and can be useful in separating
the two
samples. At Lep 2 energies the K\"orner--Schierholz--Willrodt angle $\Phi_{KSW}$
seems to be the most effective variable for this purpose, while the angle
between the two least energetic jets $\alpha_{34}$ is of little use, being
almost flat over the whole range for all samples.
From our results, namely if we
assume that the tree level background distributions closely resemble the 
actual behaviour of the background\footnote{Nagy and
Tr\'ocs\'anyi \cite{4jNLO}, however, find large corrections,
of the order of 100\% for the D parameter and for the acoplanarity.},
it appears that the differences between CC11
distributions and the dominant $q~\bar q~g~g$ background decrease
when NLO corrections to CC11 are included.

It should be stressed that four--jet shape variables in $WW$ events measure the
correlations between the hadronic decays of the two $W$'s and therefore it
should be explicitely checked whether existing codes, NLO calculations or parton
shower Monte Carlo programs, successfully reproduce the experimental curves.

The shape--variable distributions  at $\sqrt{s} = 500$ GeV are given in fig. 5.
The only difference with respect to the Lep 2 analysis is a smaller value for
the jet separation parameter $y_{cut} = 0.001$. In particular no minimum angle
between jets and either beam is required.
A separation based on shape--variables of $WW$ events from the background seems,
at first sight,
to be more difficult at the NLC than at Lep 2.  Only for the angle between the
two least energetic jets $\alpha_{34}$ the signal and background distribution
are significantly different. The former peaks in the backward direction
the latter is almost flat.
On the contrary,
the Bengtsson--Zerwas angle and  the K\"orner--Schierholz--Willrodt angle 
distribution from CC11 are almost indistinguishable from those generated
from the $q~\bar q~g~g$ background.
The sensitivity of the  Nachtmann--Reiter angle is similar at the two energies.
The signal distribution peaks at small angles, particularly at higher energies,
while the background is flatter, with a large tail which extends to
$180^\circ$. 
The distributions obtained in nQCD are closer to the full NLO results than at
lower energies.

\section*{Conclusions}
We have described the complete calculation of QCD radiative corrections 
to the process $e^+ e^- \to u~\bar d~s~\bar c$ which are
essential in order to obtain theoretical predictions for $W$--pair production
with per mil accuracy. The amplitudes we have derived are completely
differential, and realistic cuts can be imposed on the parton level
structure of the observed events.
We have presented the distribution of the average reconstructed $W$--mass
and the distribution of several four--jet shape variables at Lep 2 and NLC
energies.
The so called naive--QCD implementation of NLO corrections fails in both
instances.

\vfill\eject

\newpage
\thispagestyle{empty}
\centerline{
\epsfig{figure=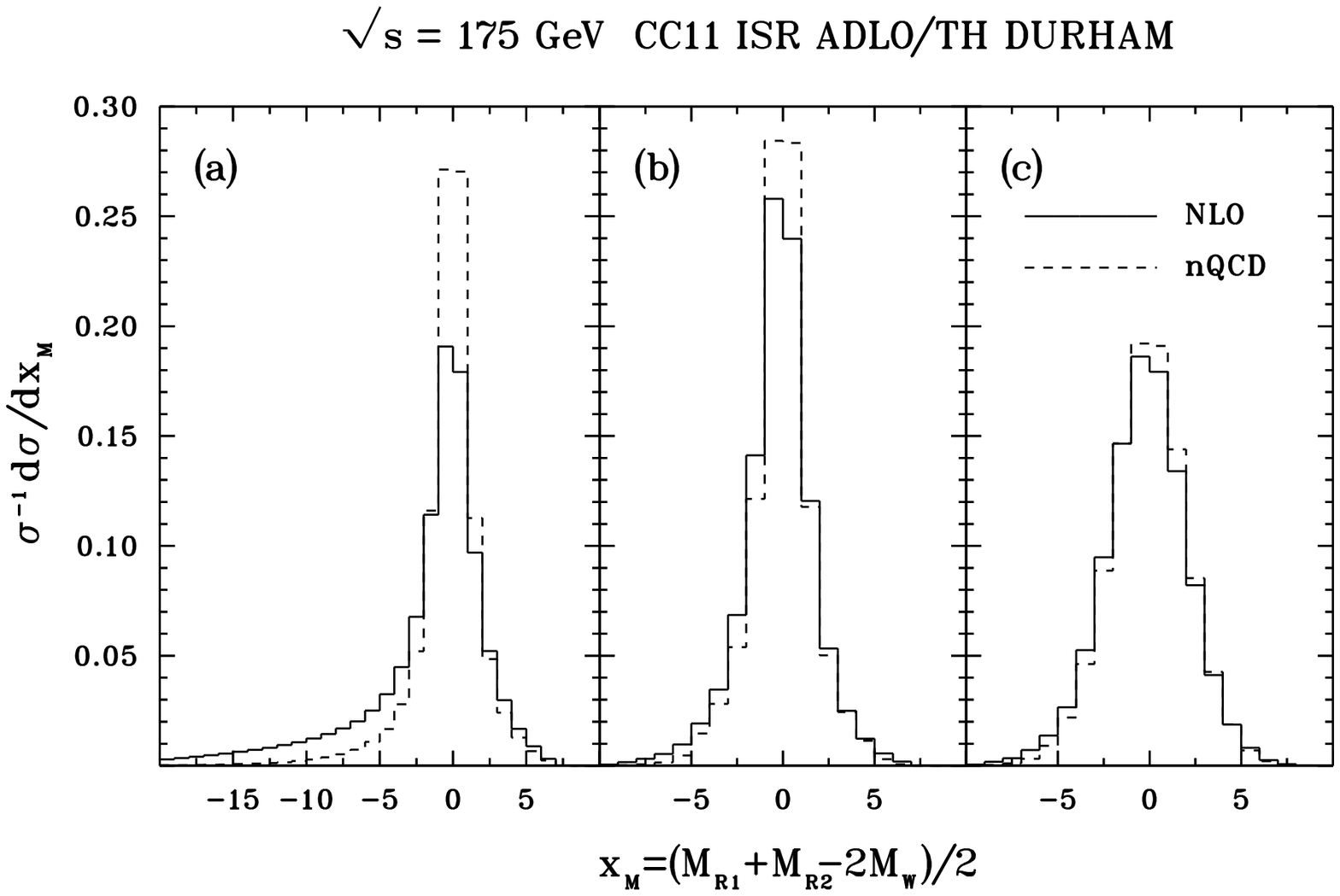,height=22cm}
}
\vspace{-6.7 cm}
\noindent
Fig. 2: Average mass distribution  at $\sqrt{s} =  175$ GeV.
All ADLO/TH cuts are applied.
The continuous histogram is the exact NLO result while the dashed histogram 
refers to nQCD. The corresponding cross sections can be found in table I.

\newpage
\thispagestyle{empty}
\centerline{
\epsfig{figure=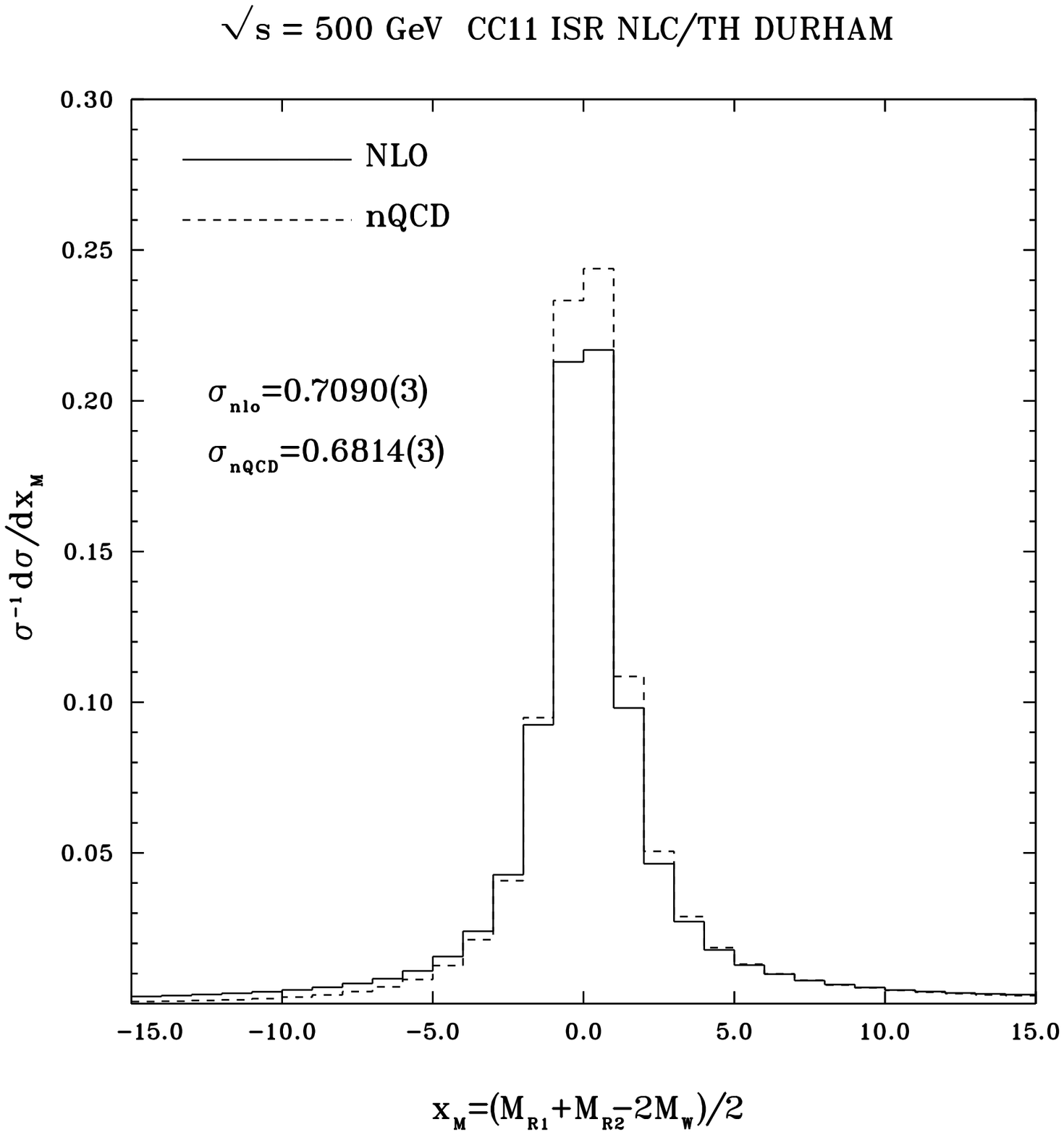,height=21cm}
}
\noindent
Fig. 3: Average mass distribution  at $\sqrt{s} =  500$ GeV.
All NLC/TH cuts are applied.
The continuous histogram is the exact NLO result while the dashed histogram 
refers to nQCD.

\newpage
\thispagestyle{empty}
\centerline{
\epsfig{figure=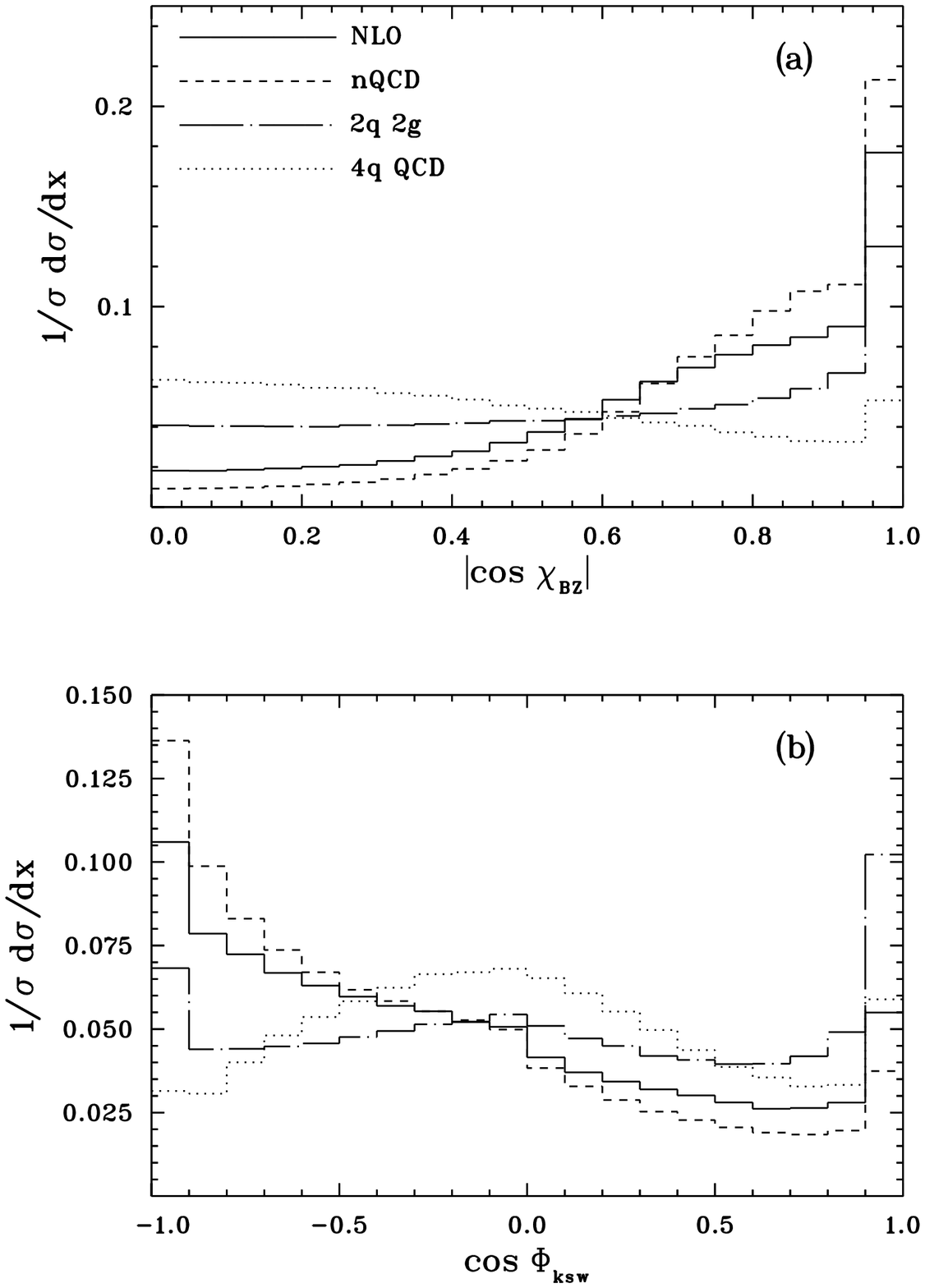,width=16cm}
}
\vspace{-1.9 cm}
\noindent
Fig. 4: 
The full NLO results (continous line) is compared with
the nQCD prediction (dashed line) and with the tree level background
distributions from $q~\bar q~g~g$ (chain--dotted line)
and $ q_1~\bar {q_1}~q_2~\bar {q_2}$ (dotted line).

\newpage
\thispagestyle{empty}
\centerline{
\epsfig{figure=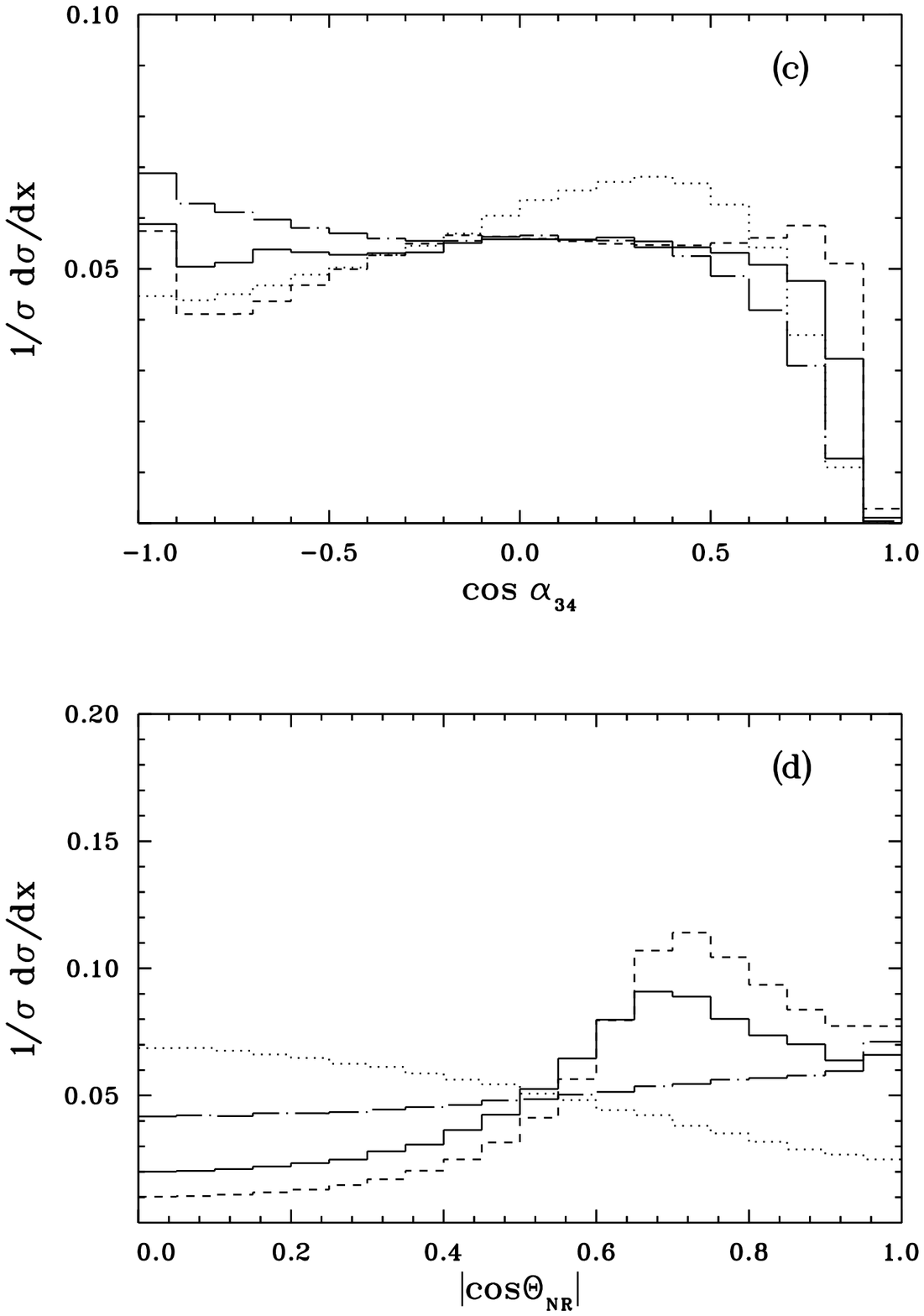,width=16cm}
}
\vspace{-1.9 cm}
\noindent
Fig. 4: Four--jet shape variables at $\sqrt{s} =  175$ GeV, continued.

\newpage
\thispagestyle{empty}
\centerline{
\epsfig{figure=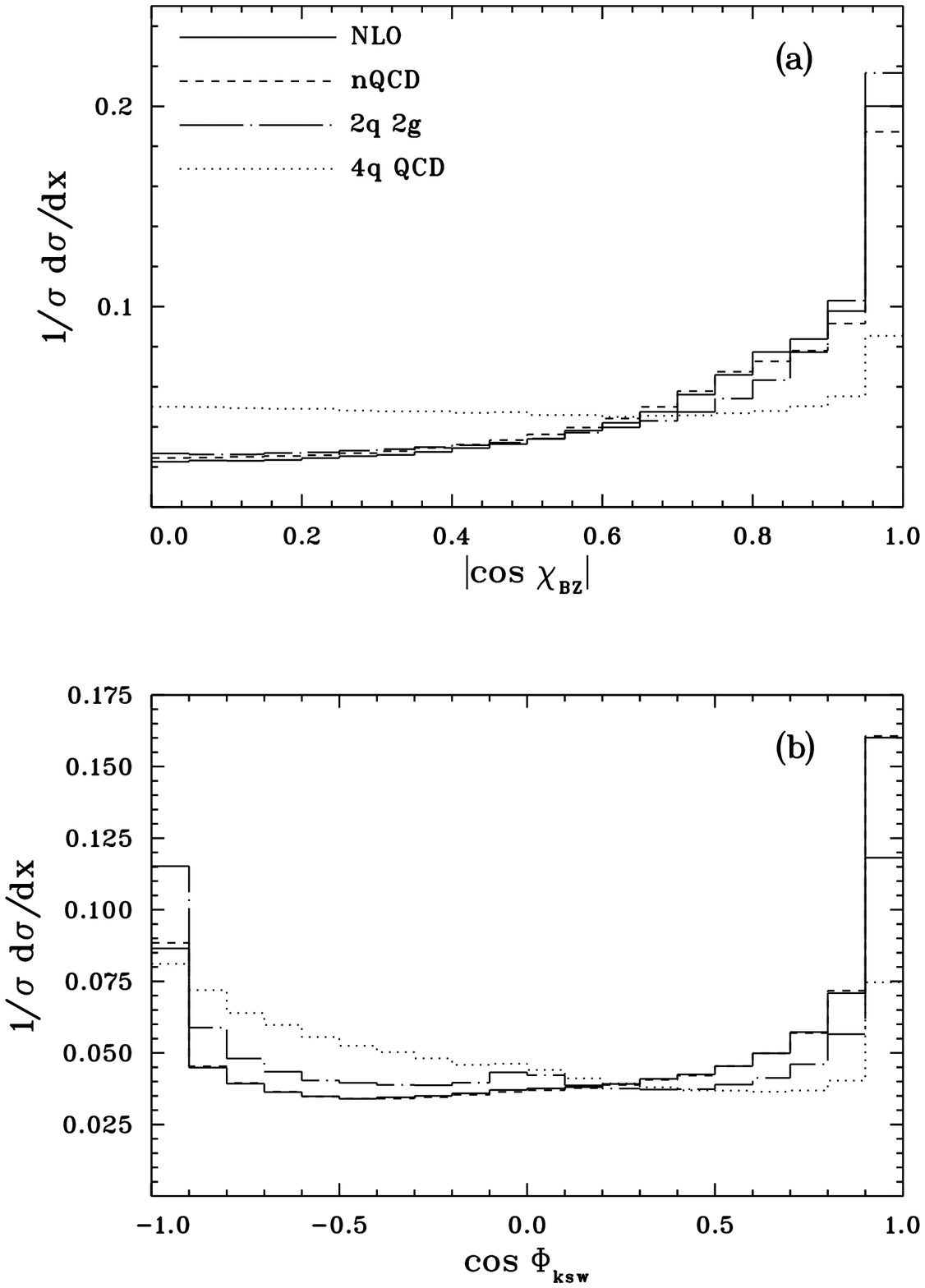,width=16cm}
}
\vspace{-1.9 cm}
\noindent
Fig. 5: Four--jet shape variables at $\sqrt{s} =  500$ GeV.
The full NLO results (continous line) is compared with
the nQCD prediction (dashed line) and with the tree level background
distributions from $q~\bar q~g~g$ (chain--dotted line)
and $ q_1~\bar {q_1}~q_2~\bar {q_2}$ (dotted line).

\newpage
\thispagestyle{empty}
\centerline{
\epsfig{figure=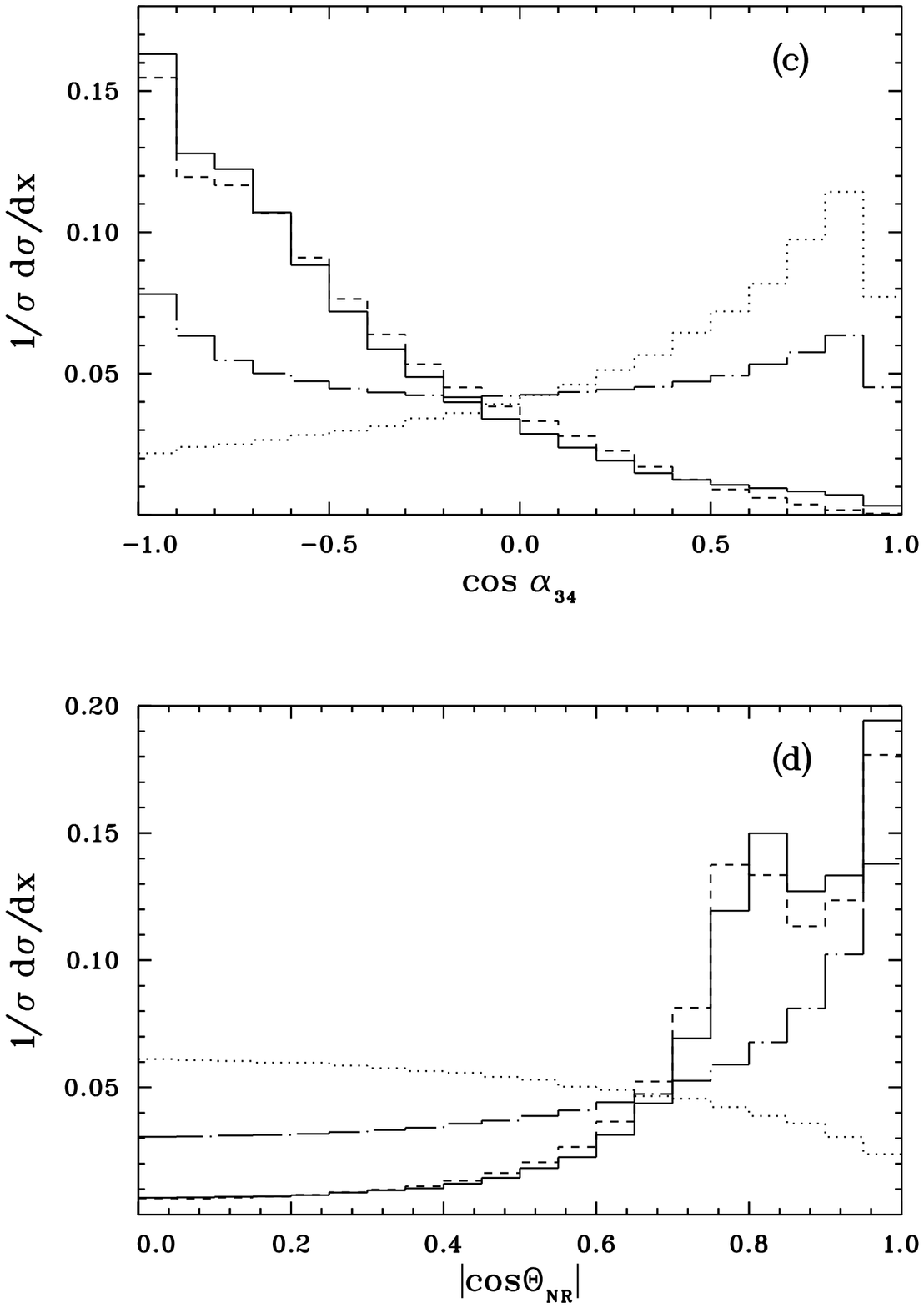,width=16cm}
}
\vspace{-1.9 cm}
\noindent
Fig. 5: Four--jet shape variables at $\sqrt{s} =  500$ GeV, continued.

\end{document}